# Nanotechnology for biosensors: A Review

Aishwaryadev Banerjee[1] (aishwaryadev.banerjee@utah.edu), Swagata Maity[2] and Carlos H. Mastrangelo[1]

[1]Department of Electrical & Computer Engineering, University of Utah, Salt Lake City, USA

[2]Department of Condensed Matter Physics and Materials Sciences, S.N. Bose National Centre for Basic Sciences, Kolkata, India

**Abstract**

Biosensors are essential tools which have been traditionally used to monitor environmental pollution, detect the presence of toxic elements and biohazardous bacteria or virus in organic matter and biomolecules for clinical diagnostics. In the last couple of decades, the scientific community has witnessed their widespread application in the fields of military, health care, industrial process control, environmental monitoring, food-quality control, and microbiology. Biosensor technology has greatly evolved from the *in vitro* studies based on the biosensing ability of organic beings to the highly sophisticated world of nanofabrication enabled miniaturized biosensors. The incorporation of nanotechnology in the vast field of biosensing has led to the development of novel sensors and sensing mechanisms, as well as an increase in the sensitivity and performance of the existing biosensors. Additionally, the nanoscale dimension further assists the development of sensors for rapid and simple detection *in vivo* as well as the ability to probe single-biomolecules and obtain critical information for their detection and analysis. However, the major drawbacks of this include, but are not limited to potential toxicities associated with the unavoidable release of nanoparticles into the environment, miniaturization induced unreliability, lack of automation, and difficulty of integrating the nanostructured-based biosensors as well as unreliable transduction signals from these devices. Although the field of biosensors is vast, we intend to explore various nanotechnology enabled biosensors as part of this review article and provide a brief description of their fundamental working principles and potential applications.

## 1. Introduction

The concept of biosensing is deeply embedded within most organic life-forms and from an evolutionary point of view, this has enabled them to survive harsh environments and predators. Common examples of this include augmented olfactory abilities of canines, the electro-sensitive nature of sharks and toxin sensing capabilities of certain algae [1][2][3]. Conventional bio-sensing techniques basically involved *in vitro* studies based on mimicking various sensory systems of such animals and according to the International Union of Pure and Applied Chemistry (IUPAC), a biosensor can be defined as "a device that uses specific biochemical reactions mediated by isolated enzymes, immune systems, tissues, organelles or whole cells to detect chemical compounds usually by electrical, thermal or optical signals" [4]. Figure 1 shows a schematic representation of the fundamental working principles of a biosensor. Clark and Lyons developed one of the first biosensors using an enzyme-electrode in the early 1960's [5]. Here, glucose was the target analyte and glucose oxidase enzyme triggered its oxidation reaction. This reaction resulted in a glucose dependent current, which when accurately measured, gave a direct indication of the glucose concentration within the sample of interest. Based on similar principles, Guilbault and Montalvo developed a potentiometric urea-sensor which utilized an enzyme-coated electrode for urea-concentration dependent voltage measurements [6]. After these preliminary efforts, the first major advancement in biosensing technology came in the form of biosensors which featured auxiliary enzymes and/or co-reactants which were immobilized with the analyte converting enzyme in order to improve and enhance the sensor output [7]. This was followed by the development of enzyme-linked immunosorbent assay (ELISA) and Surface-Plasmon-Resonant (SPR) based biosensing techniques [8][9]. However, these sensors displayed limited ability for rapid sensing and discrimination of small amounts of toxic agents embedded within large amounts of a chemically inert but complex backgrounds. Following this, the scientific community witnessed a revolutionary era of micro/nano-technology, which allowed us to explore hitherto unchartered territories and exploit the fundamentally novel features of nanoscale science and propel the field of biosensors to new heights.

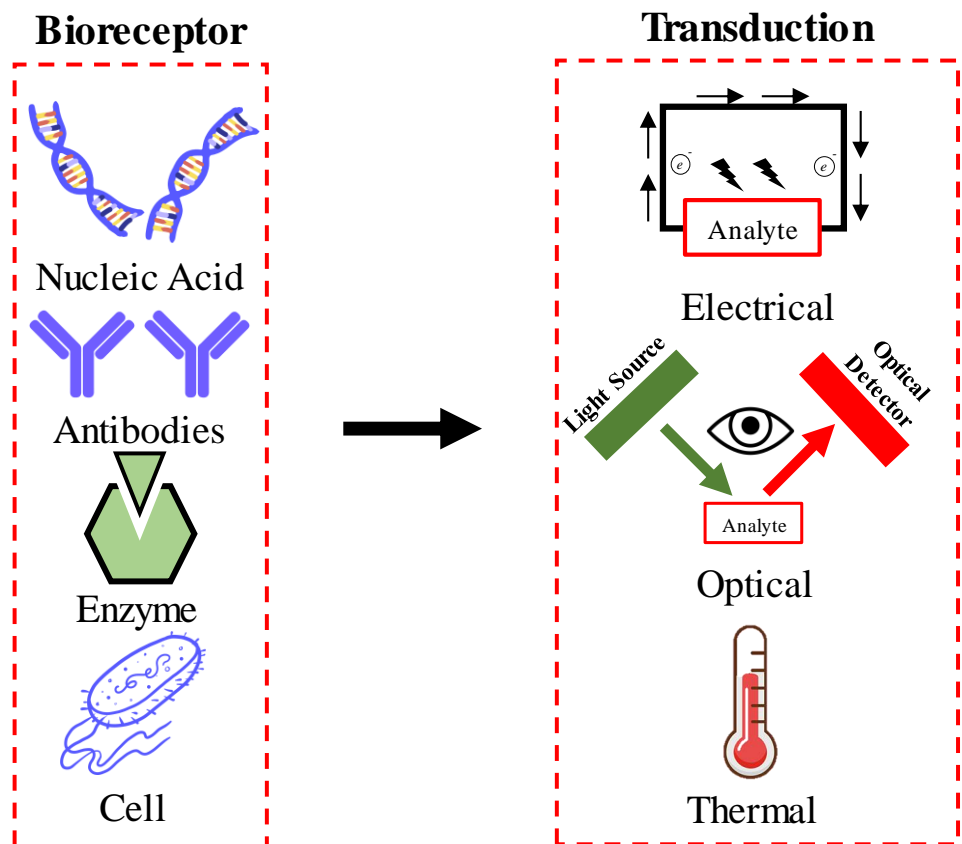

Fig. 1 : Schematic showing the working principle of biosensors based on different methods of transduction.

Advances in nanotechnology have allowed us to build structures or devices in the nano-regime, such as nanoparticles, nanotubes, nanorods and nanowires, which can directly probe and interact with the very bio-molecules we intend to detect using biosensors. Such devices display unique properties, such as excellent electrical conductivity, tunable optical, electrical and magnetic properties and show great promise for faster response and higher sensitivity at the device in comparison with conventional biosensors. Therefore, they can be exploited for a variety of bioengineering applications ranging from biosensors to drug-delivery. Nanotechnology has had a significant impact on the field of disease detection, especially by using electrochemical, electromechanical, resonance, thermal, magnetic and optical techniques. Nano-biosensors have advanced the ability to identify specific analytes and obtain detailed information regarding biomolecular profiles of various diseases. Accurate sensing and analysis of specific biomarkers has led to the development of advanced biosensor systems which are capable of sensing localized micro-environmental fluctuations, thus providing an indication of the disease, disease progression and therapeutic assessment. However, in spite of their exciting properties, these nanobiosensors are plagued with certain disadvantages such as drift, fouling, non-specificity, displaying irreproducible and non-uniform transduction signals. The following sections of this review article will provide an overview of existing nanobiosensors along with their fundamental working principles and certain applications of such devices in varied biological fields. The article begins with a very brief discussion of various types of nanobiosensors and their applications. Following this, we provide a description and applications of of thin-film based biosensors. Next, a highly detailed description of nanostructure based biosensors is also provided. In this section, we mainly describe sensors comprising of gold nanoparticles, carbon nanotubes, graphene and quantum dots. A plethora of sensor applications is also discussed with literature references. Finally, we provide a discussion of the impact nanobiosensors have had on the field of cancer research and low-power sensors for Internet of Things, including future prospects of this technology.

2. **Nanobiosensors-an overview**

Nanobiosensors are a class of sensors which are used to observe, measure and analyze biological events using sensors which have been built using the techniques of nanotechnology. They are mostly built using various forms of quantum dots, nanoparticles, nanowires and nanofilms. Most naturally occurring biological systems such as viruses, membranes, and protein complexes and their interactions take place in the nanometer regime. This makes devices whose dimensions are in the nanoscale, ideal candidates for biomedical and bioanalytical applications to obtain higher degrees of sensitivity, specificity, and faster response times compared to the traditional methods of biosensing. These include a multitude of varied applications such as using amperometric nanodevices for enzymatic detection of glucose, quantum dots as fluorescence agents for the detection of binding and even using bio-conjugated nanomaterials for specific biomolecular detection. For example, colloidal nanoparticles have been conjugated with anti-bodies for highly specific immunosensing purposes. Similarly, the electronic and optical properties of metal nanoparticles have been exploited for DNA/RNA detection and analysis. Nanomaterials/structures such as quantum dots, nanoparticles and nanotubes are the key components of any nanobiosensor system. Based on these structures, devices such as nanosensors, nanoprobes and other miniaturized systems have revolutionized the field of chemical and bio-sensing. Such nanodevices have been specifically designed to exhibit high response times and ultra-low power requirements. Nanomaterials such as metal nanoparticles, oxide nanoparticles, magnetic nanomaterials, carbon materials, Quantum Dots and metallophthalocyanines have been used to improve the electrochemical signals of biocatalytic events that occur at the electrode/electrolyte interface. Functionalized nanoparticles that are bound to organic molecules have been developed for use in biosensors. The synthesis of such nanostructured materials and nano-devices involves a varied number of techniques and the selection of a synthesis technique depends on the material of interest or the type of nanomaterial such as 0D, 1D, 2D, their sizes, and the desired quantity [10][11][12]. For example, physical techniques such as high-energy ball mixing/melt mixing, physical vapor deposition (PVD), laser ablation, electric arc and sputtering have been extensively used to develop nanostructured materials for biosensing purposes. Chemical methods of synthesizing nanostructures, such as the sol-gel process and inverse micelles formation have also been widely used. Bottom-up and top-down are the main approaches for synthesis of nanostructures materials. In the bottom-up technique, the miniaturization of material constituents followed by self-assembly results in creation of nanostructures. Such methods have been widely used for formation of quantum dots and nanoparticles from colloidal dispersions. These techniques are preferred on account of lesser defects and a more homogenous chemical composition. Contrary to this, top-down approaches involve extremely controlled processing of macroscopic structures which can be carefully engineered to build the desired nanostructure. Examples of this include IC etching techniques, ball milling and severe plastic deformation. However, a major drawback of these methods is the presence of significantly large quantities of imperfections in the surface structure. Naomaterial such as carbon nanotubes (CNTs), graphene, quantum dots (QDs), nanoparticles (NPs), and nanocomposites, have been widely investigated by the research community and used for diagnostics and biosensors in the last decade. However, the advancement of such technologies has also posed certain uncomfortable questions, for example those related to the safety of various nanomaterials, which need to be answered and addressed before a majority of these technologies can be made available for pedestrian use [13]. As shown in Figure 2, there are numerous fields of research which can be pursued while studying nanobiosensors. However, we will be focusing mainly on the fabrication, materials, characterization and direct applications of nanotechnology enabled biosensors.

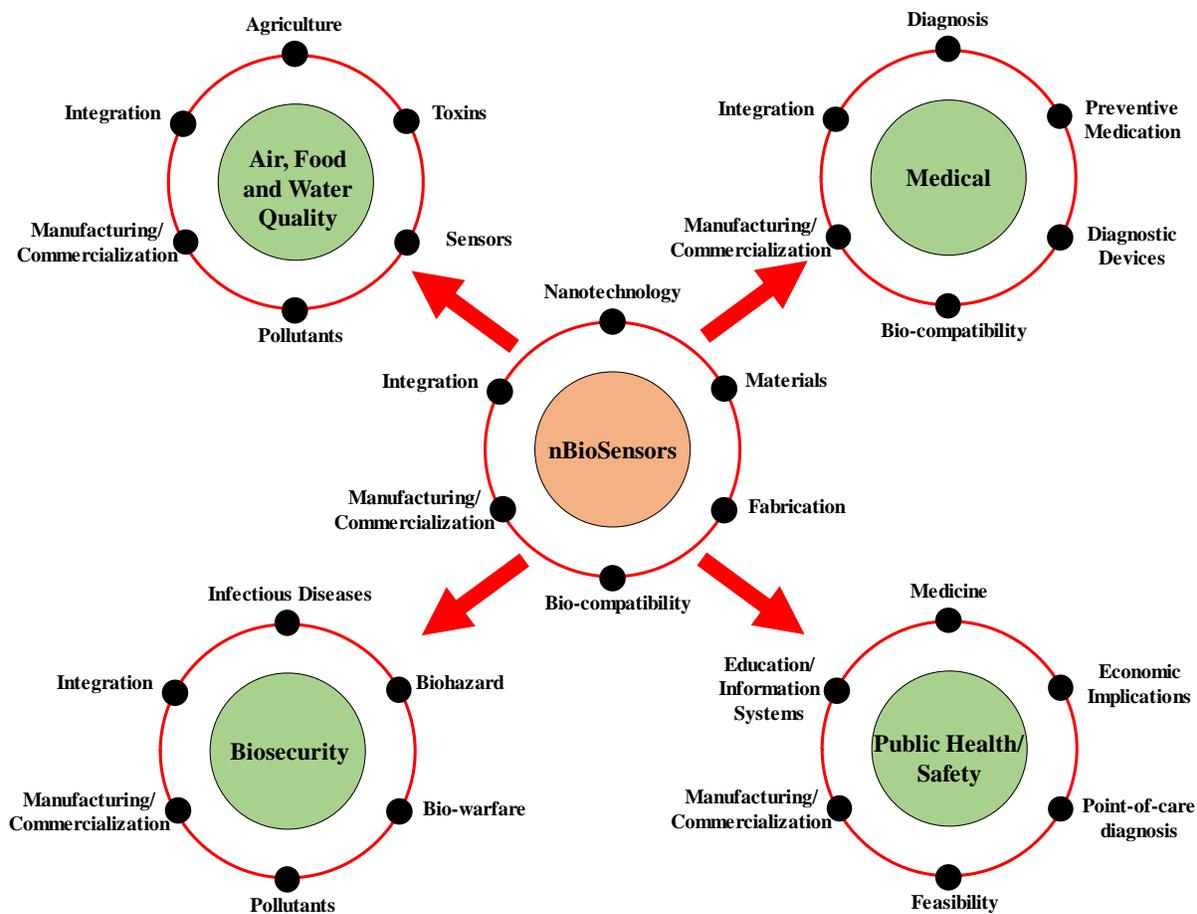

Fig. 2: Research aspects of nano-biosensors.

## 3. Thin-film biosensors

The working principle of thin-film based biosensors is based on the selective adsorption of analyte molecules on a functionalized thin-film. These biosensors act as physicochemical (optical, mechanical, magnetic, or electrical) transducers, which convert the signal resulting from the recognition of the biological analyte into another form of a measurable signal. A plethora of thin-film based biosensors can be found in literature references. These are comprised of complex structures of thin films, which give enormous functionalities to the sensors. The thin films are made from either organic and/or inorganic materials, such as metals, glass, polymers, silicon, or metal oxides. Specific structures of biological molecules can also be used as thin films. The most crucial part of such biosensors is the analyte-sensitive layer. i.e. the layer which reacts to the biomolecule. On this surface, atomic interactions, surface free energies and forces are different from the bulk of the material, and the analyte-specific reaction starts here. Surface activation can be necessary to immobilize biological analytes on sensitive layer. For this purposes, self-assembled monolayers (SAM) enabled surface modification techniques are the most widely used [14][15]. Thin films can be used to functionalize (e.g., control hydrophobicity, bioaffinity, biocompatibility, and electrical activity) the surface of biosensors with or without using surface treatment processes. For example, Parylene C, which is a polymer commonly used in the packaging of integrated circuits, is a hydrophobic polymer with a water contact angle of 87° and its surface can be made hydrophilic with oxygen plasma treatment, which allows the realization of a plethora of microfabricated microfluidic devices that function on the basis of efficient capillary flow. Other examples of similar biocompatible thin films include dielectric coatings such as $SiO_2$, $TiO_2$, $Al_2O_3$, $Si_3N_4$. These layers are crucial to the practical deployment of thin-film based biosensors and drug delivery systems. Such layers are often exploited for their superior electrical, optical, magnetic and mechanical properties in comparison with their bulk-counterparts. For example, Sokolov et al. demonstrated that a PDMS thin-film utilized as a miniature pressure sensor displayed a recovery time which was ~50 times faster than a pressure sensor which comprised of a bulk PDMS layer [16]. The most commonly used polymers for biosensing purposes include PDMS, Parylene-C, perfluoropolyether, polyetheretherketone, polypropylene and polystyrene. Besides dielectrics, metallic thin-films are widely used in biosensors to form either bond-pads for electrical probing of the analyte or as activation surfaces for appropriate functionalization. The most widely used metal for biosensors is gold due to it being biocompatible, a noble metal and its excellent electrical properties. Functionalization of Au with self-assembled thiol molecules is the most common technique for activating the gold surface. Additionally, a thin layer of metal is often deposited for improving adhesion between two layers which are naturally incompatible. Metals such as Cr and Ti have been widely used to improve adhesion between the electrically active layers (such as Au) and the underlying $Si/SiO_2$ substrates. Functionalization using organic thin-films is an equally important aspect of thin-film based biosensor technology. The functionalization process essentially consists of depositing a thin layer of organic films such as a variety of thiol molecules, proteins and nucleic acid molecules. The selection of the appropriate surface chemistry is crucial to the uniform formation of the organic films and depends on the specific material to be activated. Silane molecules are often used to functionalize inorganic surfaces, such as glass. Numerous examples of similar functionalization techniques can be found in these literature references. Figure 3 shows the schematic of the working principle of thin-film based biosensors.

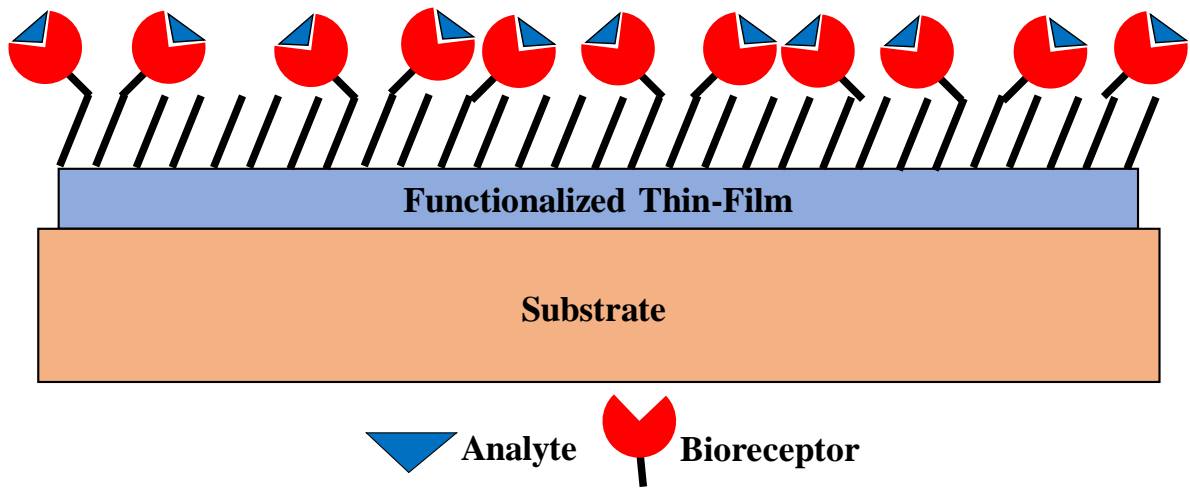

Fig. 3: Working principle of thin-film based biosensors. The thin-films are functionalized with appropriate bioreceptors to realize specific capture of analyte molecules. This capture is then measured to indirectly quantify the analyte concentration.

## 4. Nanostructures for biosensor applications

Nanostructures are attractive options for biosensor applications due to their observable quantum effects and large surface area. Exploiting these unique properties potentially enables us to design biosensors with improved characteristics compared to their traditional counterparts. Essentially, such nanostructures can be realized in 0D (for example quantum dots or nanoparticles), 1D (for example nanowires or carbon nanotubes) or 2D (for example graphene). The following sections will provide a broad overview of the most widely used nanostructures for biosensing purposes.

### 4.1 Gold Nanoparticles

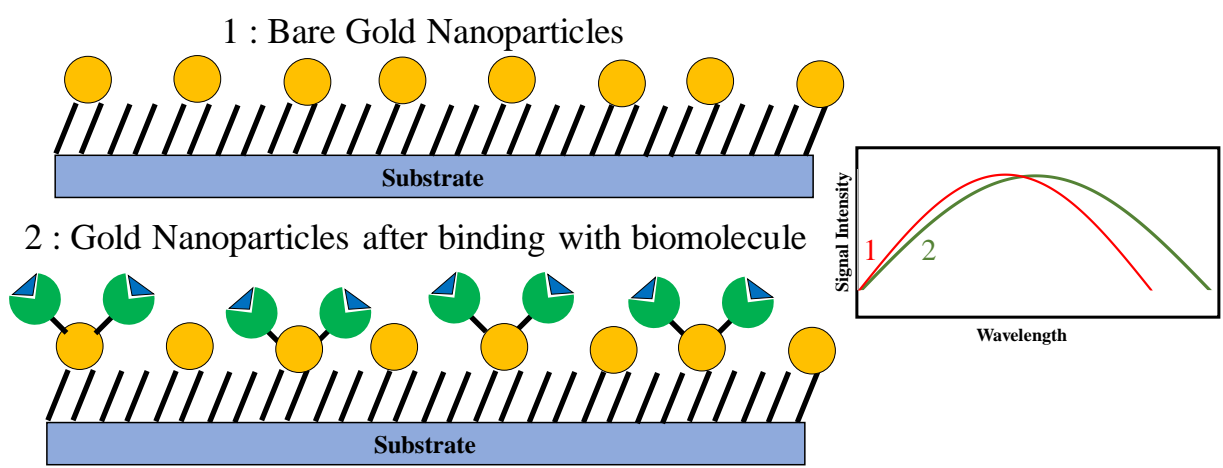

Fig. 4: Gold nanoparticle (AuNP) based biosensors. Functionalized AuNPs aid in capture of analyte molecules. The electrical/optical properties of AuNPs are utilized to obtain enhanced signals which are proportional to the analyte concentration.

The recent advances in nanotechnology have allowed for cheap and rapid synthesis of nanoparticles (NPs) and these NPs have demonstrated augmented sensitivity against certain analytes and have therefore been extensively used for biosensing purposes. For example, metallic NPs such as gold and silver nanoparticles, dielectric NPs such as $SiO_2$ or $MnO_2$ nanoparticles, polymeric NPs and semiconductor NPs such as CdS nanoparticles have been integrated with microstructures for efficient and sensitive biosensing [17]–[23]. Owing to their varying electrical characteristics, these different categories of NPs are suitable for varied roles in different biosensing systems. For example, metallic NPs are generally used for electrical probing of biomolecules. Dielectric NPs have been often employed to immobilize biomolecules and semiconductor nanoparticles have been traditionally used as labels or biomarkers. Amongst a plethora of nanoparticles which are currently used for biosensing, gold NPs (AuNPs) are the most widely used as the analyte-sensitive layer for biosensor applications due to their biocompatibility, unique opto-electronic properties, and their relatively simple fabrication and modification techniques. The size of chemically synthesized AuNPs typically lie in the range of 1-100 nm and these NPs have a high surface-to-volume ratio. AuNPs have a much higher surface energy than most other NPs and therefore are used to immobilize a wide variety of biomolecules. Due to its excellent conductive properties, AuNPs can promote very fast and direct electron transfer between a wide range of electroactive species and electrode materials. Furthermore, useful characteristics of AuNPs such as light-scattering properties and large enhancement ability of local EM fields can be used as signal amplification tags. In a nutshell, all these unique properties of AuNPs have been exploited to enhance the performance of optical, electrochemical and piezoelectric biosensors. Figure 4 shows a schematic representation of the working principle of AuNP based biosensor. The following sections briefly describe the application of AuNPs in the fields of these biosensors with relevant examples.

*AuNP based optical biosensors*: As the name suggests, these devices are designed to realize a measurable transduction between the presence or change in concentration of the analyte of interest and the optical properties of the biosensor. The fundamental principle behind gold nanoparticle assisted biosensing is the collective oscillation of plasmons in response to incident electromagnetic (EM) waves. An example of such a phenomenon is Surface Plasmon Resonance (SPR) where the interaction between the incident EM waves and the conduction band electrons of Au, is used for probing surface characteristics of the same. This interaction results in the resonant coherent oscillations of the surface conduction electrons of the metal and the characteristics of these oscillations is a function of the concentration of the analyte of interest, present on the surface. Generally, the specific binding of bio-molecules on the surface of metals results in a change of dielectric constant of the surrounding medium which then causes a change in optical characteristics of any EM wave which is incident on the metallic surface. This change is proportional to the concentration of the coated bio-molecules and therefore, accurate measurements of the reflected EM wave provide relevant information regarding the concentration of the analyte of interest. Englebienne was one of the first researchers to report a red-shift in the SPR resonant frequency of gold particles coated by a monoclonal antibody [24]. For the last two decades, significant efforts have been made to correlate the plasmon absorbance characteristics of AuNPs with the refractive index of the surrounding medium (which in turn depends on the concentration of the coated bio-molecule). The most significant advantage of AuNPs in comparison with traditional metallic thin-film devices is that, AuNP based sensors display augmented SPR signals which results in higher sensitivity of the biosensor and accuracy of measurement as well as a lower limit of detection.
Some significant examples include the work of Lin et al. who used the SPR technique and demonstrated an optical-fiber based biosensor, sensitive to organophosphorous pesticides [25]. The presence of the pesticide led to an increase in the local refractive index which led to change in the attenuation of the incident light. Therefore, measuring the attenuation of the light could easily determine the concentration of the pesticide. The presence of AuNPs significantly improved the sensitivity of the bio-sensor. Using similar principles, He et al. used AuNPs for specific detection of DNA hybridization [26]. Generally, literature suggests that an increase in sensitivity of about 2-3 orders of magnitude is observed due to the presence of AuNPs as compared to other unamplified biosensing techniques. Li et al. performed SPR imaging to analyze single nucleotide polymorphisms in genomic DNA [27]. The common underlying principle in the aforementioned examples is of AuNP assisted signal amplification. This amplification is essentially the result of electronic coupling between the localized surface plasmon of AuNPs and the propagating plasmon on the SPR gold surface, and the increase of effective mass of the immobilized analytes, due to the high density and high molecular weight of AuNPs. Additional examples of AuNP based amplified biosensors include the contributions of Okamoto, Matsui and Qi [28]–[30]. Furthermore, AuNPs have also been integrated with other photonic structures. For example, Tseng et al. developed an immuno-sensor based on AuNPs integrated with an optical fiber, conjugated with recognition proteins, where the shift in interference fringe was proportional to the analyte concentration [31]. Finally, AuNPs have also been used for Surface Enhanced Raman Scattering techniques. For example, Cao et al. demonstrated sensitive detection of oligonucleotide-targets by functionalizing oligonucleotides and Raman-active dyes on AuNPs [32].

*AuNP based electrochemical biosensors*: Electrochemical biosensors are devices which are designed specifically for transduction of bio-chemical signals into measurable electrical signals. These sensors have been widely investigated in the last couple of decades due to their relatively simple fabrication process and low-cost detection capabilities. As mentioned above, the excellent electrical properties, biocompatibility and catalytic properties of AuNPs make them an attractive component of electrochemical sensors. Integrating AuNPs with existing electrochemical sensors leads to enhanced transduction signals due to an "amplified" electrode surface and augmented electron transfer between redox centers in proteins. AuNPs have also been used as catalysts for electrochemical reactions. The role of AuNPs can be broadly classified as (1) electron transfer mediators and (2) immobilization platforms. Usually, the active center of most bio-molecules (for example a protein) is surrounded by a thick layer of electrically non-conducting shells, which impedes flow of electrons between the electrically active center and the probing electrode, which leads to a poor output electrical signal. Natan et al. demonstrated for the first time, the amplification of electrical signals due to augmented electron transfer between the active center and the electrode [33]. Similarly, Willner et al. studied the electronic properties of AuNP coated with glucose oxidase [17]. The enzyme coated electrodes displayed fast electron transfer between the redox center of the protein and the AuNPs. AuNPs have also been integrated with polymer based electrochemical biosensors. Such biosensors which employ AuNP dispersed in various polymers have exhibited augmented stability and reusability. For example, Xu et al. have demonstrated biosensing properties of a composite material comprising of AuNPs dispersed in carboxymethyl chitosan for hydrogen peroxide sensing using electrochemical methods [34]. Ju et al. built a disposable immunosensor, capable of multi-analyte testing, which included four different horseradish peroxidase-labeled antibodies immobilized on AuNPs, on screen-printed carbon electrodes with chitosan sol-gel [35]. AuNPs are ideal for immobilizing bio-molecules, in contrast to their bulk counterparts since bio-molecule adsorption on bulk materials have often lead to degradation of their properties. For example, adsorption of protein bio-molecules on the surface of bulk materials have often led to denaturation. Adsorption of such bio-molecules on AuNP surfaces have shown no signs of degradation due to the biocompatibility and high surface-energy of gold. Additionally, due to their inherently higher surface area than their bulk counterparts, AuNPs can immobilize far greater number of protein molecules, which therefore allows realization of more sensitive bio-sensor devices. For example, Andreescu et al. utilized immobilized periplasmic glucose receptors on AuNPs and measured their electrical properties as a function of glucose concentration to develop a novel glucose sensor [36]. Similarly, based on this principle of bio-molecule immobilization on AuNPs, many devices have been developed for sensitive and selective bio-sensing. Another vital application of this principle can be found in electrochemical DNA sensors which involve oligonucleotides immobilized on AuNPs. For example, using thiol end-groups, oligonucleotides have been immobilized on AuNPs by Kang et al [37]. Specifically, the thiol end-group modified oligonucleotides at the 5' phosphate end were immobilized on the AuNPs and due to the high surface-to-volume ratio, the hybridization amount of DNA was greatly enhanced and therefore easily detected. A similar technique of biosensing was employed by Fang et al. where they immobilized oligonucleotides with a thiol group at the 5'-phosphate end on AuNPs, which were later self-assembled on cystamine modified Au electrodes, for detecting DNA hybridization [38].

In addition to the above mentioned applications, AuNPs have been used in conjugation with other nanostructures such as carbon nanotubes (CNTs) and graphene to further enhance their immobilization and binding capabilities. For example, Cui et al. have built an electrochemical biosensor based on the conjugation of AuNPs and CNTs [39].

Although gold is an inert element, small-sized AuNPs have demonstrated their ability to act as a catalyst. This property of AuNPs has been widely exploited for bio-sensing purposes. Such properties of AuNPs are generally believed to be a result of quantum-mechanical effects arising due to their much augmented surface-to-volume ratio. AuNPs have therefore been effectively utilized to reduce overpotentials of crucial electrochemical reactions as well as engineer the reversibility of certain redox reactions. For example, Tuener et al. demonstrated that 1.4 nm sized AuNPs provided catalyst-like features for oxidation of styrene by dioxygen [40]. Furthermore, as the AuNP size increased to ~2 nm, the nanoparticles we no longer effective catalysts. Similarly, Raj et al. investigated into the catalytic properties of AuNPs by demonstrating their role in enhanced oxidation of NADH [41]. As an application of the catalytic property of small-sized AuNPs, Bharati et al. developed a glucose biosensor based on the AuNP mediated redox reactions of $H_2O_2$ [42]. Another similar example involves bio-sensors utilizing the catalytic properties of AuNPs and developing an enzyme-free glucose sensor by immobilizing AuNPs inside a sol-gel and on an electrode surface. Other properties of AuNPs which are exploited for the purpose of electrochemical biosensing include the ability of these nanoparticles to stabilize the biomolecules they are interacting with whilst not degrading their properties and simultaneously enhancing the stability of the biosensing devices.

*AuNP based piezoelectric biosensors*: The final category of AuNP based biosensors which will be explored in this article are AuNP based piezoelectric biosensors. In these sensors, any mass changes initiated due to biological events are detected by the quartz crystal microbalance (QCM) method, which is based on the piezoelectric effect. High density AuNPs with enhanced surface-to-volume ratios are integrated with piezoelectric biosensors to amplify the inherently weak signals. For example, Jiang et al. demonstrated a microgravimetric DNA biosensor integrated with AuNPs [43]. In this device, AuNPs functionalized with oligonucleotide probes, were immobilized on the surface of the QCM. Hybridization with the probes led to a change in mass, which was detected by the QCM. Similarly, Li et al. also developed a DNA hybridization piezoelectric biosensor [44]. Other examples of highly sensitive piezoelectric biosensors based on similar principles include the role of AuNPs as signal amplification tags for detecting DNA mutation, single-base mismatch detection and *E.Coli* detection [45], [46]. Besides DNA hybridization sensors, AuNP based piezoelectric biosensors have been used for building devices for ligand-sensing by Yu et al [47]. Literature suggests that using AuNPs reduced the limit of detection by 3 orders of magnitude. Examples of hybrid piezoelectric biosensors include a conjugation of AuNP and hydroxyapatite for surface immobilization of target analyte[48]. This hybrid nanomaterial has multi-adsorption sites and augmented solubility and dispersibility. Using this hybrid material for piezoelectric biosensing, achieved higher sensitivity against analyte molecule in comparison with devices employing only AuNPs or hydroxyapatite.

**4.2** Carbon Nanotubes

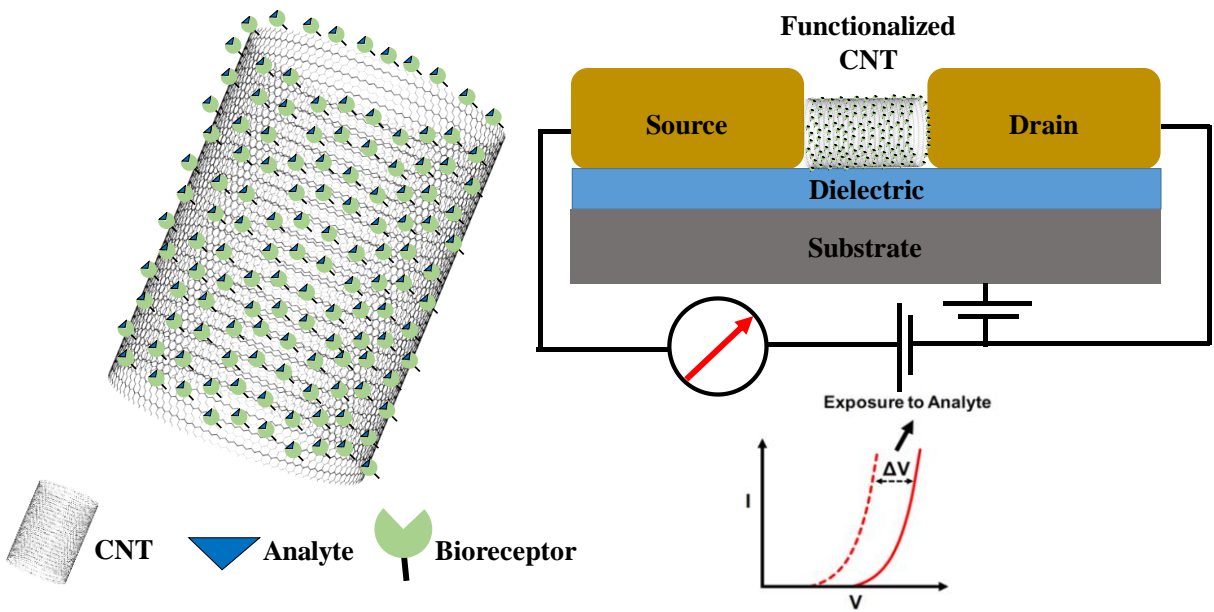

Fig. 5: Schematic of Carbon nanotube (CNT) functionalized with bioreceptors to facilitate the capture of specific analyte. Schematic of CNT-based field effect transistor. Successful capture of analyte molecule leads to shift in threshold voltage of the FET device, which is in indirect measure of analyte concentration.

Carbon nanotubes (CNTs) are cylindrical structures comprising of "rolled-up" sheets of a single-layer of carbon atoms. Depending on their structure, they can be classified as either single-walled or multi-walled. After their discovery in 1991 [49], CNTs have been the subject of significant interest and intrigue due to their unique structural, chemical and electrical properties. In the last three decades, CNTs have been employed in varied fields of research including nanoelectronics and biomedical engineering. Specifically, due to the exponential advance of nanotechnology and nano-manufacturing, CNTs have been widely used to develop novel biosensors. Similar to other nanostructures, CNTs offer a large surface/volume ratio which enables immobilization of much larger quantities of biomolecules on their surface in comparison with bulk-materials. Due to their unique properties, CNTs can be exploited to develop novel probes for a variety of biomolecules. For biosensing applications, CNTs can be used as (1) specific capture platforms, (2) probes for transduction of analyte induced electrical signals and (3) novel mechanism of in vivo probing due to the ability of CNTs to cross a biological membrane. Figure 4 shows a generalized example of a CNT based biosensing technique. Depending on the mechanism of biosensing, CNT based biosensors can be classified as electrochemical and electronic sensors, immunosensors and optical biosensors. Such devices are usually extremely sensitive and find applications in the health industry for early stage detection of biomarkers of a variety of diseases such as cancer. The following sections will briefly discuss these CNT based biosensors.

*CNT based electrochemical biosensors*: As mentioned above, electrochemical biosensors are designed for transduction of bio-chemical signals into measurable electrical signals. These sensors have been widely used due to their low-cost, ease of fabrication and use as well as small footprint. Like AuNPs, CNTs have the ability to enhance the electron transfer which make them ideal additions to electrochemical biosensors. CNTs display a high surface-to-volume ratio, augmented sensitivity, electrical conductivity and chemical stability, as well as high electrocatalytic effect, which make them ideal candidates for biosensors which require enzymatic reactions for accurate sensing. Therefore, a wide variety of CNT based electrochemical sensors have been developed in the last two decades. For example, Wang et al. developed such devices for detecting metabolites and protein biomarkers [50]. CNTs have also been used for developing sensitive glucose sensors. The structural self-alignment of glucose oxidase on electrodes using SWNTs as conduction paths between enzyme redox centers and probing electrodes have been demonstrated by Patolsky et al [51]. Using similar principles, CNT based biosensors have been developed for accurate sensing of cholesterol, using electrodes functionalized with holesterolesterase, peroxidase and oxidase [52]. Santos et al. demonstrated biosensors employing functionalized CNTs which can detect nitric oxide [53]. Similarly, Prasad et al. developed an epinephrine sensor and Kress et al. built a dopamine sensor in rats based on CNTs [54], [55]. Zelada-Guillen et al. demonstrated a novel aptamaer based sensor, which used CNTs grafted with protein-specific RNA aptamers, for detection of certain glycoproteins in blood [56]. Some additional examples of similar biosensors include Zhang et al. demonstrating the detection of cellular nitric oxide by single-stranded d(AT)15 DNA oligonucleotide curled around CNTs and Jin et al. building SWNT-based hydrogen peroxide biosensors [57], [58]. Based on the method of transduction, electrochemical biosensors can be further classified as either potentiometric, amperometric, conductometric, piezoelectric or voltametric. For example, Fei et al. developed a cysteine detector using Pt/CNT electrodes based on principles of cyclic voltammetry [59]. Antiochia et al. fabricated an amperometric PSA sensor using PSA-antibody coated on CNTs [60].

*CNT based optical biosensors*: As describe in previous sections, optical biosensors are designed to realize a measurable transduction between the presence or change in concentration of the analyte of interest and the optical properties of the biosensor. Lubbers and Oppitz were the first to develop an optical biosensor for $CO_2$, $O_2$ and alcohol detection [61]. Since then, there has been widespread interest in developing optical biosensors for detecting, investigating and quantifying biological processes in vitro and in vivo. Such biosensors can be classified based on their specific optical transduction mechanism, for example surface plasmon resonance, fluorescence, absorbance/reflectance etc. These sensors can be further classified as either probing or reacting. Probing sensors provide optical information depending on the differences in interactions between the analyte and the sensor whereas reacting sensors provide different optical responses based on biologically induced chemical reactions. CNTs have also been widely pursued as an attractive addition to the already existing bio-sensing principles to develop novel biosensor technology. Inherently, unfunctionalized CNTs have displayed useful characteristics such as low fluorescence stability, intensity, and biocompatibility. However, after being appropriately functionalized, CNTs can be engineered to display changes in fluorescent emission signals upon exposure to the intended analyte. Additionally, CNTs have been employed to detect changes in local dielectric function and also have displayed augmented resistance to unwanted photobleaching affects, which makes them attractive candidates for optical biosensor applications. For example, NIR detection of ATP living cells have been demonstrated by Kim et al. using CNT/luciferase conjugates [62]. Heller et al. have demonstrated a nitroaromatics sensor, using principles of photoluminescence in the NIR spectrum, based on CNTs functionalized with peptides [63]. CNTs functionalized with genetically altered M13 have also been used for deep tissue imaging [64]. Using similar principles, single-walled CNTs, functionalized with oligonucleotide labelled dies, have been used by Yang et al. to develop an optical biosensor for single-stranded DNA [65]. In these sensors, the fluorescence of the CNT composite quenched until the single-stranded DNA binds and released the labeled oligonucleotide from the CNTs. Polymers dispersed with CNTs have been used to optically detect different metabolites (riboflavin, L-thyroxine and oestradiol). Zhang et al developed a label-free optical sensor for the protein, troponin T, using CNTs functionalized by chitosan using NIR fluorescence techniques [66], [67].

### 4.3 Graphene

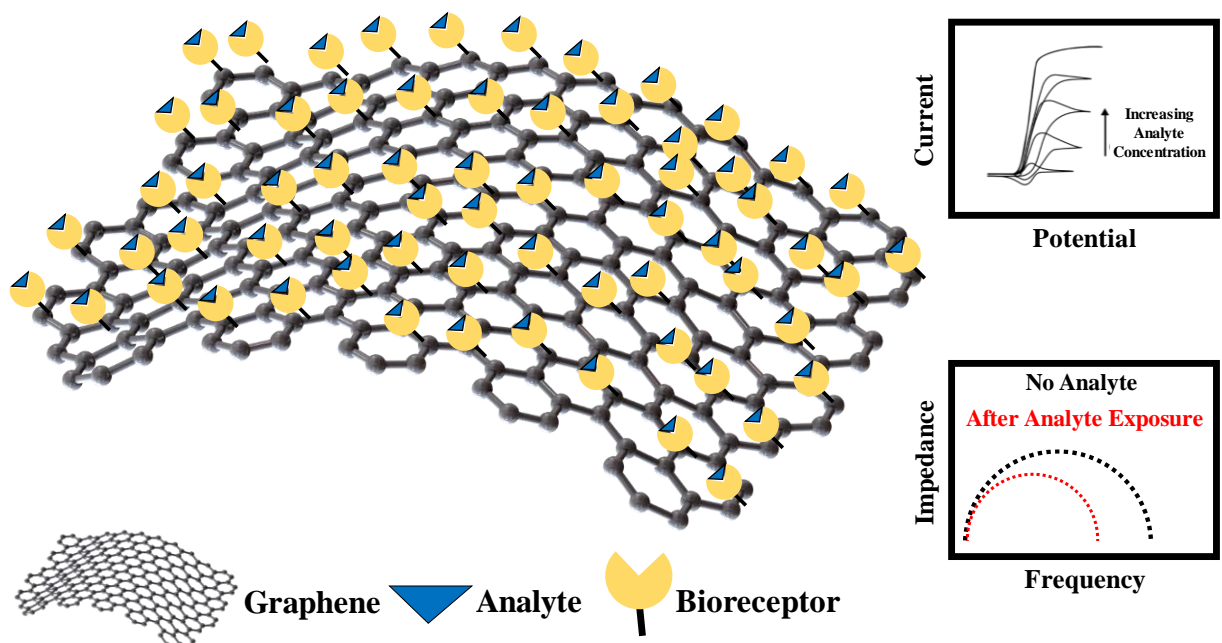

Fig. 6: Schematic showing analyte capture by appropriately functionalized graphene monolayer. The capture can be electrically measured by recording the impedance properties of the graphene layer, before and after analyte capture. Additionally, cyclic voltammetry measurements of the graphene layers display a change in I-V plots as a function of analyte concentration.

Graphene was first theoretically explored by P.R. Wallace in 1947 [68] and then later synthesized by the novel prize-winning scientist, Konstantin Novoselov and his team using a simple scotch-tape dispenser [69]. Graphene and is a carbon structure with $sp^2$ hybridization in two dimensions and can be viewed as

"carbon-sheets". These sheets can then either be stacked horizontally to form a 3D graphite structure or can be simply rolled to form nanotubes. Due to the π-conjugation present in the graphene structure, it exhibits excellent thermal, electrical and mechanical properties. Graphene can be synthesized using a variety of techniques, including chemical vapor deposition (CVD), exfoliation of graphite, liquid phase exfoliation of graphite, reduction of graphene oxide, surface segregation and molecular beam epitaxy. However, most of these are low-throughput techniques and unfortunately, there has been limited advancement in repeatable and thorough batch-fabrication of graphene. Using techniques such as atomic-force microscopy (AFM), scanning tunneling microscopy (STM) and Raman Spectroscopy the most vital properties of graphene have been determined as follows: the surface area of a single-layer of graphene is ~2630 $m^2g^{-1}$ and the Young's modulus is 1 TPa with an intrinsic mobility of $2\times10^5$ $cm^2v^{-1}s^{-1}$. The thermal conductivity of graphene has been determined as $5\times10^3$ $Wm^{-1}K^{-1}$. It is optically almost transparent with a measured transmittance of 97.7% and it is extremely conductive with a conductivity of ~$10^4\,\Omega^{-1}cm^{-1}$. Amongst its widespread applications, graphene has been mostly used to build high-speed electronic devices for storage and energy harvesting. It has been used to fabricate high frequency transistors, photo-detectors and biosensors for a plethora of applications such as DNA sequencing. Figure 6 shows a functionalized graphene layer and the working principle of a graphen-based electrochemical biosensor. The following sections will briefly discuss the graphene-based electrochemical biosensors.

*Graphene based electrochemical biosensors*: As mentioned above, due to its excellent electrical properties, graphene can be integrated with the electrodes of electrochemical biosensors for enhanced biosensing. In addition to graphene's augmented conductivity, it is chemically stable, has low cost and is electrochemically inert. However, in certain cases, graphene has also been used as a catalyst for numerous bio-chemical redox reactions. In such bio-sensors, enzymes are immobilized on the graphene layer, which are used for specific analyte recognition purposes. The immobilization can be realized using a variety of techniques including adsorption technique, covalent binding, electro-polymerization and layer-by-layer immobilization. Amongst the variety of applications of graphene based electrochemical biosensors, the most important are detection of heavy metals, phenols, pesticides and other pollutants such as hydrogen peroxide or microorganisms for environmental monitoring purposes. Electrochemical sensors, based on principles of potentiometry and voltammetry, have been widely pursued to detect the presence of heavy metals and are being pursued widely due to their low cost, portability, reduced response time and augmented sensitivity. Literature shows that electrochemical sensors for detecting heavy metals make use of noble metals such as Au and Pt for electrodes or different forms of carbon [70]. Specifically, carbon-based nanomaterials are effective adsorbents of heavy metals and have been often used as heavy-metal scrubbers for cleaning purposes. Therefore, it follows that such materials have widespread application in bio-sensors for heavy metal detection. Amongst all carbon-based materials, graphene is the most affective in adsorbing organic and inorganic pollutants. For example, Kong et al. demonstrated a graphene-based electrochemical biosensor for detecting low concentrations of $Cu^{2+}$ (1.5 nM) and $Pb^{2+}$ (0.4 nM), based on principles of square waver voltammetry analysis techniques [71]. In this device the gold electrode of the sensor was functionalized with aryl diazonium salt followed by the immobilization of covalently bonded graphene sheets. The use of inorganic nanoparticles, such as salts, is required to prevent the formation of graphene agglomerates due to its hydrophobic nature. Hybrid polymer/graphene materials have also been used to detect $Cu^{2+}$ and $Pb^{2+}$ to further lower the limit of detection. Zhou et al. developed a bio-sensor for detecting $Pb^{2+}$ using graphene-quantum dot structures. These modified quantum dots were fabricated from graphene oxide, which were powdered into smaller pieces and chemically reduced using the green's technique with polystyrene sulfonate and L-ascorbic acid [72]. Another such example is where an array of AuNPs were immobilized over graphene sheets for detection of $Hg^{2+}$. Another highly sensitive and selective technique of detecting $Hg^{2+}$ is by using graphene functionalized with aptamers [73]. The limit of detection was determined to be ~ 10 pM. Graphene and AuNP composites have also been utilized for the specific detection of Cr (VI) [74]. Graphene has also been utilized for detection of another commonly found toxic material – phenols. Although phenols, such as Bisphenol A, are electroactive compounds, the electrochemical response is arduous to measure. The electrical properties of graphene have been employed as catalysts for these electrochemical reactions. Graphene sheets which have been negatively doped have been used for sensitive detection of Bisphenol A [75], [76]. Another interesting approach of low limit of phenol detection was presented when graphene was functionalized with enzymes immobilized with glutaraldehyde [77]. Recently, stacked layers of graphene in the form of μ-pillars, immobilized with enzymes were used to amperometrically detect 50 nM of phenols [78]. Some other techniques frequently used to amplify the detection capabilities of phenols include modification of graphene with β-cyclodextrin for low-limit detection (0.09 μM) of 2-chlorophenol and 3-chlorophenol. The nanocavities of β-cyclodextrin help in magnifying the surface-to-volume ratio and improve transduction capabilities and output signal [79]. Literature suggests that electrostatic interactions between enzymes and graphene generally lead to a heightened adsorption of enzyme on the graphene surface and graphene's electrocatalytic properties helps improve the analytical response of such bio-sensors. For example, acetylcholinesterase immobilized on graphene has been used to detect the presence of pesticides such as organophosphates [80]. The principle behind the detection of these molecules is that organophosphates generally inhibit the enzymes which are immobilized on the graphene sheets. This results in an "oxidation-current" which is a measurement of the concentration of organophosphate molecules. Polymers such as Nafion have been used for improved enzyme immobilization on graphene. For example, a combination of Nafion and graphene-oxide have been used to form a nanocomposite which is used to modify the biosensor electrode followed by enzyme immobilization for detection of organophosphates such as dichlorvos [81]. Another similar biosensor was developed which included organophosphorus hydrolase enzyme, immobilized in Nafion for detection of paraoxon. The fabricated device displayed excellent electrochemical properties, high sensitivity, fast response and low limit of detection [82]. Graphene has also been integrated with metallic nanoparticles to develop sensitive biosensors for detection of organophosphate molecules. For example, graphene sheets with AuNPs and poly-(diallyldimethylammonium chloride) immobilized with the enzyme acetylcholinesterase, allows for much greater enzyme activity and this phenomenon is used for highly sensitive detection of organophosphate molecules [83]. Addition of metallic nanoparticles to graphene-based biosensors, provides much greater electrocatalytic activity thereby allowing for much lower limit of detection. In addition to metallic nanoparticles, CNTs have also been employed (in combination with dispersants like Polyaniline) to reinforce such biosensors since they have the ability to aid in electron transfer between analyte molecules and the probing electrodes, which helps in the realization of sensitive detection of the analyte. Other biosensors used for similar purposes include non-enzymatic biosensors which comprise of a hybrid stack of graphene and CNT for detecting organophosphates such as methyl parathion [84]. Another example of such an electrochemical biosensor is based on Cobalt (III) oxide (CoO)-reduced graphene oxide, which was developed for the purposes of detecting carbofuran and carbaryl in agricultural produce [85]. Generally, non-enzymatic graphene biosensors have displayed greater stability in sensor response, as compared to the enzyme-assisted based sensing of pesticides such as organophosphates. Besides heavy metals and pesticides, graphene based biosensors have been used to detect the presence of hydrogen peroxide and certain microorganisms. Zhou et al. developed a biosensor which comprised of hybrid-multilayers of graphene and chitosan with microperoxidase salt as a probing electrode [86]. This configuration allowed for low limit of detection (2.5 μM) of $H_2O_2$. Song et al. reported a similar device which comprised of a multilayer electrode, consisting of molybdenum-disulfide–graphene and horseradish peroxidase for ultra-low limit of detection of $H_2O_2$ [87]. *Escherichia coli* (*E. coli*) has been previously detected using graphene based bio-sensors using functionalized graphene layers. Literature suggests that probing electrodes which comprise of a hybrid graphene and CNT layer as well as the presence of graphene oxide provides a higher probability of *E. Coli* capture compared to bare graphene [88]. In other configurations, specific detection of bacteria is enabled by functionalizing the graphene layer with antibodies specific to the bacteria intended to be captured [89]. Another type of biosensor for *E. Coli* detection, which does not involve the use of antibodies is based on using the adsorption of the slightly negatively charged *E. Coli*. on the surface of the p-type graphene.

Another bacterium which has been similarly detected (using either antibodies or aptamers) using graphene based biosensors is *Staphylococcus aureus* [90].

### 4.4 Quantum Dots

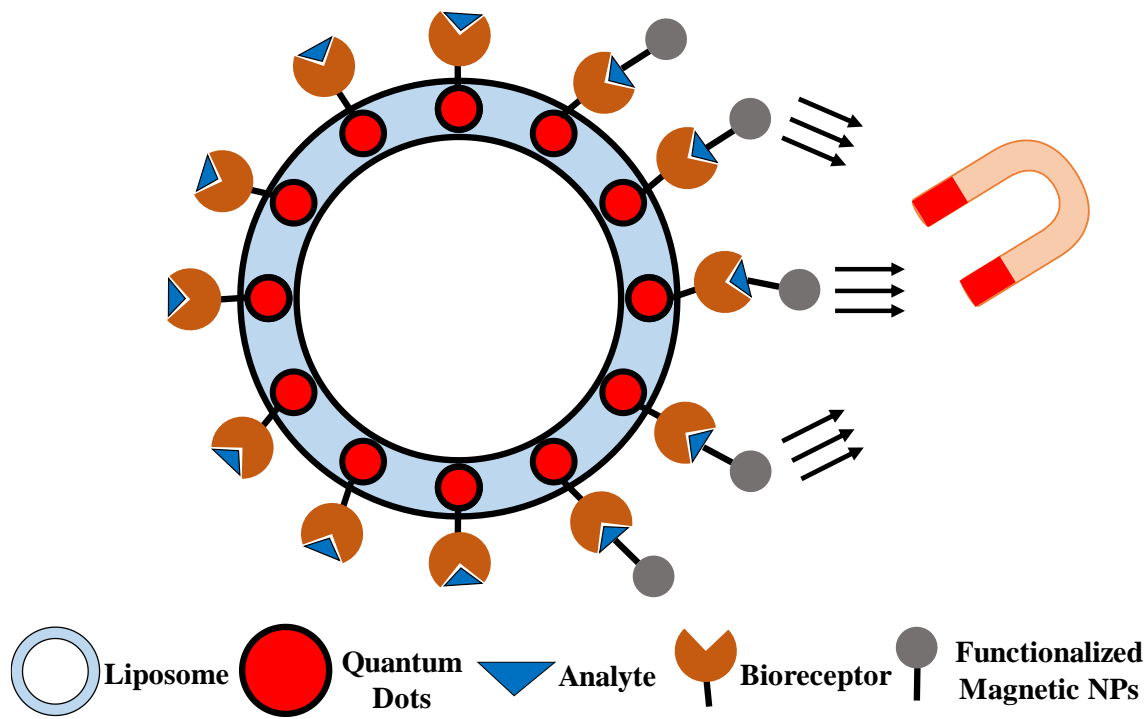

Fig. 7: Schematic shows quantum-dot (QD) based biosensors. The QDs are integrated with liposomes and functionalized with appropriate bioreceptor. They facilitate capture of specific analyte, thereby forming a sandwich hybrid with a magnetic capture-probe at one end.

Quantum dots are colloidal nanostructures mostly built using semiconductor materials belonging to the groups of II–VI, III–V, or IV–VI in the modern periodic table. Due to quantum mechanical effects, quantum dots display opto-electrical properties which are significantly different that conventional bulk materials. These properties make quantum-dots attractive candidates for bio-sensing purposes. Alivisatos and Nie were the first to develop a quantum-dot biosensor and imaging system based on fluorescence, in 1998 [91], [92]. Since then, quantum-dots have been widely used for biosensing purposes. The fluorescent properties of quantum-dots can be tuned simply by altering their geometrical measurements, and therefore, using them is advantageous in comparison with traditional fluorescent materials such as proteins or organic dyes. Additionally, quantum-dots display enhanced brightness and stability against bleaching compared to traditional fluorescent materials. These nanostructures can also be easily functionalized with a plethora of biomolecules such as antibodies, peptides and DNA to realize modified quantum-dot probes which interact with other biomolecules with an augmented level of specificity. Quantum-dots have also been used in combination with traditional metallic probing electrodes since they display photo-electrochemical properties. For example, quantum-dots which have been immobilized on electrodes display enhanced electrical activity and electron transfer between themselves and the electrode, when exposed to certain wavelengths of light. These properties of quantum-dots have led to widespread research and development in the field of quantum-dot based biosensors. Based on their sensing mechanism, we can classify these biosensors as (1) fluorescent (2) bioluminescent (3) chemi-luminescent and (4) photoelectromechanical. Figure 7 is a schematic of the working principle of a quantum-dot based biosensor.

*Quantum-dot based fluorescent biosensors*: As mentioned above, due to its high sensitivity and enhanced stability as well as abilities to measure different analytes simultaneously, have led to significant development of quantum-dot based fluorescent biosensors. Fluorescence based resonant energy transfer is the most widely used technique for quantum-dot based biosensors. The advantage of the resonant technique is that it enables detection of target biomolecules without any requirement of washing and separation steps. Additionally, high SNR (signal-to-noise ratio) and highly sensitive techniques such as single molecule detection methods have been used for quantum dot based biosensors for the detection of biomolecules such as DNAs, DNA methylation and numerous enzymes. For example, Zhang et al. developed a DNA assay where liposomes encapsulated hundreds of quantum-dots to form hybrid complexes to enable binding of target DNA. This technique was used to accurately and simultaneously detect genetic sequences of HIV-1 and HIV-2 at $10^{-18}$ molar concentrations [93]. Shamsipur et al. demonstrated a quantum-dot based biosensor using resonant energy transfer techniques for nano-molar concentration detections of *Helicobacter pylori-specific* DNA [94]. Krull et al. developed a point-of-care rapid diagnostic paper-based nucleic acid assay using quantum-dots with fluorescent resonant energy transfer techniques [95]. This device exhibited highly specific and sensitive (~fmol concentration) detection of Cy3-DNA hybridization. Zhang et al. demonstrated sensitive and high SNR detection of DNA methylation using a single quantum-dot based biosensor [96]. Using fluorescence based resonant energy transfer techniques, Hildebrandt and Qiu demonstrated quantum-dot based simultaneous detection of three different microRNAs: hsa-miR-20a-5p, hsa-miR-20b-5p and hsa-miR-21-5p [97]. Zhang et al. improved on this work, buy demonstrating multiplexed biosensing using a single quantum-dot based biosensor by employing fluorescence resonant energy transfer techniques [98]. Ho et al. developed a quantum-dot based fluorescent biosensor for sensitive detection of miR-141 in prostate cancer, with a detection limit of 1 pM [99]. Wang et al. developed a quantum-dot based fluorescent biosensor for detection of DNA and microRNA [100]. Sang and Wang et al. fabricated a quantum-dot based fluorescent biosensor for accurate detection of ten different glycoproteins using fluorescence polarization techniques with a limit of detection of 0.15 μM [101]. Hildebrandt et al. built a quantum-dot based immunosensor for highly sensitive (1.6 ng/mL) detection of prostate-specific antigens in 50 μL serum samples, for prostate cancer diagnosis [102]. Based on the quenching property of $H_2O_2$, Tang et al. demonstrated a paper-based biosensor for accurate detection of carcinoembryonic antigen with a limit-of-detection of ~6.7 pg/mL [103]. Using similar principles, Zhang et al. developed a quantum-dot based biosensor for simultaneously detecting multiple proteins by using a protein-binding dye bromophenol blue [104]. Kim et al. built a quantum-dot based immunosensor for enzyme-less and ultra-sensitive detection (<$10^{-18}$ molar concentration) of the protein, myoglobin [105]. Tyrakowski and Snee fabricated an antibody-free quantum-dot based highly sensitive biosensor (with a limit of detection ~ 3.06 pmol/mL) for sensitive detection of streptavidin [106]. Li et al. demonstrated a fluorescent based caspase assay based on the principles of inner-filter effect of mobilized AuNPs immobilized on quantum-dot fluorescence and achieved detection in the range of pM concentration [107]. Zhang et al. fabricated a quantum-dot based biosensor for sensitive detection of DNA glycosylase activity with human 8-oxoguanine-DNA glycosylase 1. Zhang et al. developed a single quantum-dot based fluorescent biosensor for highly sensitive detection of DNA methyltransferase [108]. The same group fabricated a single quantum-dot based biosensor for rapid and highly sensitive detection of terminal deoxynucleotidyl transferase activity [108]. Petryayeva and Algar built a quantum-dot based fluorescent biosensor for sensitive detection of three different proteolytic enzymes (trypsin, chymotrypsin, and enterokinase), with a limit of detection in the range of nM concentrations [109]. Using fluorescent techniques, Ngeontae et al. fabricated a quantum-dot based biosensor for detecting abnormal levels of Adenosine-5'-triphosphate (ATP) [110].

*Quantum-dot based bioluminescent biosensors*: A major bottleneck to the practical implementation of fluorescent biosensors based on quantum-dots is the inherent requirement of appropriate illumination sources. Additionally, such sensors suffer from high background noise signals due to direct excitation of the acceptor fluorophore. This can be solved by using bioluminescence resonance energy transfer technique where a bioluminescent luciferase acts as the energy donor which acts as the catalyst for the oxidation of the substrate which generates the emission light and which can then be transferred to the acceptor to complete the bioluminescence resonance energy transfer process. For example, Shen et al. demonstrated sensitive and selective detection of enrofloxacin using a quantum-dot based bioluminescent biosensor [111]. Additionally, Jin and Tsuboi fabricated an optical biosensor for the detection of apoptosis cells using quantum-dot biosensors [112].

*Quantum-dot based chemiluminescent biosensors*: These devices are based on the principle of emission of light due to a chemical reaction. Quantum-dots are exploited for their catalytic properties which can enhance redox chemiluminescent reactions and thereby improve the transduction signal. Therefore, quantum-dot based chemiluminescent biosensors have been widely developed for their low background noise and high sensitivity. For example, Ju et al. built a quantum-dot based chemiluminescent biosensor for sensitive detection of carcinoembryonic antigen with a limit of detection in the range of 24 fg/mL [113]. Based on similar principles, Yao et al. demonstrated sensitive biosensing of single stranded DNA binding protein [114]. Similarly, Zou et al. fabricated an electrogenerated chemiluminescence biosensor, capable of detecting dopamine present in trace amounts in urine and CSF fluid samples [115].

*Quantum-dot based photoelectrochemical biosensors*: In these biosensors the analyte dependent photoexcitation is the source of transduction energy, which is measured for sensing purposes. Such devices are generally easy and cheap to build, and are designed to be portable for practical purposes. They have been used to build highly sensitive sensors for detecting DNAs and proteins. For example, Ju et al. developed photoelectrochemical biosensors based on quantum-dots, functionalized with ZnO nano-sheets for highly sensitive detection (~0.93 fM concentration) of DNA [116]. Zhu et al. developed similar DNA assays with enhanced sensitivity by using CdTe quantum-dots [117]. The limit of detection was measured to be as low as approximately 27 aM. Such biosensors have also been used by Chen et al. to detect extremely low-levels of cardiac troponin T ($1\times10^{-7}$ g/L). Chen used CdS quantum-dots with $TiO_2$ to realize modified ITO electrodes [117]. Finally, Dai et al. recently fabricated a device for ultra-sensitive detection of carcinoembryonic antigen using CdTe quantum-dots [118].

**4.5** Other nanomaterials used for purposes of biosensing

*Chitosan based biosensors*: Chitosan is a naturally occurring biodegradable polymer which has been widely used for biosensing purposes due to its biocompatibility, nontoxicity and high film stability. For example, Wang et al. developed an amperometric biosensor using a polymer matrix of graphene oxide/chitosan/hemoglobin nanocomposite for highly sensitive detection of nitromethane, where the limit of detection was ~ 5μM [119]. Prabhakar et al. fabricated a biosensor which comprised of a hybrid layer of chitosan and $Fe_2O_3$ nanocomposite films for detection of malathion [120]. CNTs functionalized with chitosan/glucose-oxidase immobilized on polypyrrole/Nafion was used as glucose sensors [121]. Some other notable examples of the application of Chitosan on biosensors can be found in these literature references [122]–[143].

*Dendrimers based biosensors*: Dendrimers are three-dimensional gel based matrices which can essentially be considered as monodispersed macromolecules which are structurally constructed around a core unit. Dendrimers can be suitably engineered to alter their properties such as structural uniformity, monodispersity and globular shape to realize highly efficient and sensitive biosensors. Literature suggests that integrating dendrimers with existing biosensor technologies leads to sensors with augmented sensitivity, specificity, enhanced stability and repeatability. Dendrimers have therefore been used to realize novel forms of electrochemical, fluorescent and gravimetric biosensors. These macromolecules have also been used to enhance the transduction capabilities of biosensing techniques such as surface enhanced Raman scattering (SERS) and surface plasmon resonance (SPR). A plethora of dendrimer based biosensor applications can be found in these literature references. [144]–[160].

## 5. Some specific applications of nanobiosensors

In the previous sections, the article provides an extensive overview of biosensors comprising of nano-films/structures and their many applications. In this section, we provide a brief review of the application of such biosensors specifically in two extremely important fields of research (1) cancer diagnostics and (2) low-power, portable sensing techniques for Internet of Things (IoT) applications.

### 5.1 Cancer diagnosis

Due to its high mortality rate, cancer has been the subject of widespread and incessant research over the globe. On account of their enhanced sensitivity, selectivity, ease-of-use and superior analytical performance for rapid sensing, nanostructure-based biosensors are highly attractive candidates for cancer detection. The following sections will provide a brief literature survey of nanostructure-based biosensors and bio-sensing techniques which employ nanostructures, for cancer diagnosis.

*CNT based biosensors*: Feng et al. developed a paper-based bipolar electrode, functionalized with multi-wall CNTs for sensitive electrochemiluminescent detection of prostate specific antigen(PSA) [161]. Baj-Rossi et al. demonstrated CNT based biosensors for detection of drugs aimed at treatment of breast cancer[162]. Ovádeková et al. developed a screen-printed carbon electrode, modified by CNTs and AuNPs for detection of berberine, an isoquinoline plant alkaloid which demonstrates significant antimicrobial and anti-cancer activity [163]. Liu et al. fabricated a biosensor which comprised of tricosane-functionalized single-walled CNTs for sensitive detection of VOCs present in the breath, which are a potential indication of lung-cancer [164]. Park et al. built a CNT based biosensor for detection of the cancer marker galactin-3 [165]. Zheng et al. demonstrated electrochemical detection of HeLa and HL60 cancer cells using folic acid-functionalized polydopamine-coated CNTs [166]. Fayazfar et al. built a sensitive label-free detection platform for sensing TP53 gene mutation [167]. Shobha et al. demonstrated early-stage detection of prostate cancer using CNTs functionalized with specific DNA strands [168]. Lerner et al. fabricated a CNT based field-effect transistor, immunosensor for detection of osteopontin, which is a biomarker of prostate cancer [169]. Abdolahad et al. used vertically aligned CNTs for detection of SW48 cells from grade IV human colon tumors [170]. Liu et al. developed a multilayer CNT based biosensor for detection of liver cancer cells [171]. Veetil and Ye et al. developed CNT based immunosensors for probing cancer biomarkers [172]. Malhotra et al. demonstrated ultrasensitive electrochemical immunosensing based on CNTs for detecting very low concentrations of Interleukin-6 (Il-6) [173]. Wan et al. demonstrated simultaneous detection of PSA and Interleukin-8 (Il-8) using screen-printed carbon electrodes [174]. Arkan et al. fabricated a biosensor for detection of HER2 in breast cancer patients, which comprised of AuNP immobilized on CNT liquid electrodes [175]. Zerda et al. developed optical CNT biosensors for detection of alpha(v) beta(3) integrins [176]. Wang et al. developed a highly sensitive fluorescent assay for detecting ultra-low concentrations of cyclin A peptide, which is overexpressed in certain human cancers [177].

*Quantum-dot based biosensors*: Ho and Wilner et al. demonstrated a quantum-dot based biosensor for the detection of miR-141 in prostate cancer [99]. Zhang et al. developed a quantum-dot biosensor for rapid detection of terminal deoxynucleotidyl transferase (TdT), which is a biomarker for leukemic disease [178]. Kim et al. demonstrated fluorescence imaging of metalloprotein in cancer cells using quantum-dot based fluorescent probes [179]. Jie et al. demonstrated highly sensitive cancer cell detection (98 cells/mL) using an amplified electrochemiluminescent biosensors employing magnetic quantum-dots [180].

*Surface plasmon resonance based biosensors*: Surface-plasmon resonance (SPR) based biosensors have been actively used for detecting cancer biomarkers in serum samples [181]. Using similar techniques, sensitive detection of carcinoembryonic antigen has also been reported [182]. Carcinoembryonic antigen is associated with colorectal and lung cancers. Other examples of SPR based biosensor applications for cancer cell detection include ultra-sensitive detection of prostate-specific antigen, tumor markers, human chronic gonadotropin, leukocyte cell adhesion molecules, breast cancers and oral cancers [183]–[186]. Prasad et al. demonstrated SPR based ultrasensitive bio-sensing of tumor necrosis factor alpha (TNF-α) at concentrations in the range of fmol [187]. Another example is the use of RNA aptamer microarrays and sensitive SPR techniques for detection of vascular endothelial growth factor (VEGF) (at 1 pM concentration), a biomarker for lung cancer, breast cancer and colorectal cancer [188]. Jang et al. demonstrated highly sensitive detection of a prostate specific antigen (PSA) using optical fiber SPR techniques which employed the use of sandwich assay and analyte specific antibodies [189]. Similar techniques have also been used for detection of oral cancer biomarkers. Another method of ultrasensitive detection of PSA was demonstrated by the use of super-paramagnetic particles with biomarker-specific antibodies and SPR techniques [190]. The limit of detection was determined to be 10 fg/mL. Similar biosensors have also been used for sensitive detection of pituitary hormones [191].

*Quartz crystal microbalance (QCM) based biosensors*: QCM based biosensors integrated with nanoparticles have been used for sensitive detection of PSA and PSA–alpha 1-antichymotrypsin biomarkers, with a limit of detection measured in the range of 0.29 ng/mL [192]. QCM techniques have also been used for enhancing the output signals of electrochemical impedance spectroscopy sensors by employing DNA aptamers against PSA by using thiol-mediated surface chemistry modification of Au surface on the sensor [193]. Wang et al. used QCM techniques for detecting single-nucleotide polymorphism in the p53 tumor suppressor gene [194]. QCM biosensors have been used with AuNPs to demonstrate highly sensitive detection of hybridization of DNA fragments of the p53 gene near codon 248 [195]. Similar techniques have also been employed to detect the p16 gene. Poly (2-hydroxyethyl methacrylate) (PHEMA) nanoparticle assisted QCM techniques have also been used for sensitive and accurate detection of breast cancer cells including MCF7 and MDA-MB 231 [196].

*Magnetic nanoparticles based biosensors and drug-delivery*: Functionalized magnetic nanoparticles are vital for noninvasive and highly sensitive detection of cancer cells. Kumar et al. utilized magnetic nanoparticles, functionalized with target-specific peptides for enhanced tumor uptake and more pronounced silencing effects [197]. These magnetic nanoparticles are generally iron-based magnetic oxides due to their properties of biocompatibility, saturation magnetization and resistance against oxidation processes. Extensive details of such nanoparticles and their applications in the field of cancer research can be found in these literature references [198]–[207]. Furthermore, such nanoparticles have been further modified to also act as efficient drug delivery systems. For example, paclitaxel, which is an anticancer drug, has pronounced side-effects including inflammation of the veins and tissue damage. Paclitaxel has been labeled with magnetic nanoparticles to avoid the above-mentioned side-effects [203].

5.2 Low-power sensors for Internet of Things

With the meteoric rise of the Internet of Things (IoT) in the last two decades, humanity is on the verge of witnessing the advent of a new digital nervous system for the world. With the practical deployment of an interconnected network of cameras, microphones, and a plethora of sensor and actuator systems, we seem to be inching towards a truly 'smart' society. The earliest realization of the IoT was in 1982 when a group of local programmers at Carnegie Melon University connected a Coca Cola machine to the internet to check if there was a drink available in the machine and if it was cold. With the term "Internet of Things" being coined in 1999, the practical implications of this technology have been widespread in the last twenty years, and it has evolved immensely. By the year 2013, the IoT became an amalgamation of various technologies, including the internet, wireless communication, GPS, embedded systems, and MEMS devices. The latest advances in the IoT technology include development of smart-wearable electronics for health monitoring, smart-homes, pre-integrated IoT platforms, merging of Artificial Intelligence, and Machine Learning algorithms with stand-alone devices and implementation of low-cost and power-efficient sensor and sensor-networks.

At the backbone of the entire IoT framework lies a variety of sensors and sensor systems that collect diverse data and share it across the network of interconnected smart devices. This makes it possible for these devices to operate autonomously and improve the effectiveness and functionality of the vast IoT network. Some of the critical sensors which are being extensively used for these purposes are temperature, proximity and pressure sensors, water and air-quality monitoring sensors and IR sensors. However, the practical realization of such an omnipresent and densely interconnected system demands the availability of low-power sensors and sensor systems for energy-efficient device operation and communication between the smart devices. For example, continuous air quality monitoring and asset tracking requires the sensors to be battery operated and deployed in remote areas of the world. Furthermore, to minimize energy consumption, most of these sensors are designed to typically operate in "sleep" or "stand-by" mode where it consumes very low power and record data only in the presence of a stimulus or according to a pre-determined operation protocol. Research intended for the development of low-power sensors and sensor-systems for IoT based applications has been actively pursued by a significant proportion of the scientific community in the last couple of years and has led to major advancements in the field of nanotechnology based sensor and sensor systems. The following section provides some key references across this field.

Chikkadi et al. demonstrated the ultra-low-power operation of self-heated, suspended carbon nanotube gas sensors [208]. Ngoc and Wang presented the design and fabrication of low-power and self-sustaining $SnO_2$ nanowire gas sensors for IoT and portable applications [209], [210]. Liu et al. demonstrated low power gas sensors fabricated using Silicon Nanowires/$TiO_2$ core-shell heterojunctions [211]. Han et al. proposed a new low-power gas sensor utilizing 1-D Si nanowire and 2-D $SnO_2$ thin films [212]. Cho demonstrated a novel low-power consuming and high-sensitivity Schottky $H_2$ sensor based on Si nanomembranes [213]. Alreshaid and Stetter discussed multiple ink-jet printed nano-sensors for IoT and smart-city applications [214], [215]. Mamun et al. describe various wearable sensors for environmental monitoring for IoT applications [216]. Long et al. demonstrate low-power gas sensing using 3D porous nanostructured metal-oxide sensors for application related to the IoT [217]. Weiss provides an elaborate discussion of low-power and highly-sensitive magnetic sensors [218]. Villani et al. describe a novel self-sustaining ultra-low non-invasive voltage/current sensor for IoT based solutions [219]. Kassal et al. demonstrate the development and characterization of an ultra-low-power radio-frequency identification sensor tag for use in IoT based applications [220]. Laubhan et al. proposed the implementation of a low-power IoT wireless sensor network for the detection of motion, humidity and temperature [221]. Kuo et al. presented a design of a low-power and long-range sensor node for the next generation IoT platform [222]. Kamakshi et al. proposed a design for a nano-Watt powered CMOS temperature sensor for ultra-low power IoT applications [223]. Garulli et al. describe a smart temperature sensor design based on the architecture of CMOS 65 nm technology for use in the IoT network [224]. Kim et al. have recently developed a novel sensing technique for highly sensitive biosensing based on the quantum-tunneling effect using nanogap break-junctions [225]–[233]. Furthermore, Mastrangelo et al. have also been responsible for developing zero-power and highly-sensitive polymeric sensors for VOC detection [234]–[240].

IoT sensors in the field of healthcare is also an active area of research. Beach et al. presented an ultra-low-power ECG monitor integrated with the SPHERE IoT platform [241]. Xican et al. provide low-power sensor designs for IoT and mobile healthcare applications [242]. Gatouillat et al. describe the development and evaluation of an ECG based cardiorespiratory IoT sensor [243]. Medu et al. describe low-power memory technology for sensors related to IoT based wearables and portable medical devices [244].

Researchers worldwide are highly invested in developing state-of-the-art sensor-nodes/systems and system architecture catered to IoT applications. For example, Hayashikoshi and Gogoi proposed a low-power Multi-Sensor platform for IoT applications [245], [246]. Tresanchez et al. describe the design of a cost-effective and low-power embedded wireless image sensor node for IoT applications [247]. Fayyazi et al. describe an ultra-low-power memrisitive neuromorphic circuit for IoT based smart sensors [248]. Ameloot et al. provide a low-power, autonomous, compact, wireless IoT sensor node based on LoRa and SigFox technologies [248]. Ma et al. introduced a visible light-enabled indoor localization system for low-power IoT sensors [249]. Mois et al. evaluated self-powered environmental monitoring systems that use Bluetooth Low Energy (BLE) beacons that operate in the IoT environment [250]. We have also noticed a tumultuous increase in media coverage of this technology as well, where research in low power sensor technology intended for IoT based applications is being boosted by some of the major industrial giants such as Apple, Google, Amazon and On Semiconductor [251]–[260]. Statistics predict that the potential growth in this industry is exceptionally high since only 0.06% of all possible devices have been optimized for the IoT platform [261].

## 6. Conclusion

In conclusion, this review article aims at providing the reader a detailed overview of some of the most commonly used biosensors which have been realized by nanotechnology enabled thin-films and miniaturized structures. These include thin-film, gold nanoparticles, carbon nanotubes, graphene and quantum-dots. Numerous examples from literature have also been cited to provide the applications of such biosensors. Furthermore, the article also describes the influence of such biosensors specific to the vital fields of cancer research and low-power sensors for the Internet of Things including future commercial prospects of this technology.